\documentclass[11pt]{article}
\usepackage{latexsym}
\usepackage{amssymb}
\usepackage[dvips]{graphicx}
\usepackage{epsfig}
\usepackage{rotating}

\usepackage{amsfonts}
\usepackage{amsmath}
\newlength{\Taille}

\input xy
\xyoption{all}

\newcommand{\flechebas}[1]{
  \settoheight{\unitlength}{\mbox{$#1$}}
  \settowidth{\Taille}{\mbox{~${\scriptstyle #1}$}}
  \addtolength{\unitlength}{4ex}
  \begin{picture}(0,1)
    \put(0,1){\vector(0,-1){1}}
    \put(0,0.5){\makebox(0,0){${\scriptstyle #1}$ \hspace{\the\Taille}}}
  \end{picture}}
\newcommand{\flechehaut}[1]{
  \settoheight{\unitlength}{\mbox{$#1$}}
  \settowidth{\Taille}{\mbox{~${\scriptstyle #1}$}}
  \addtolength{\unitlength}{4ex}
  \begin{picture}(0,1)
    \put(0,0){\vector(0,1){1}}
    \put(0,0.5){\makebox(0,0){\hspace{\the\Taille}${\scriptstyle #1}$ }}
  \end{picture}}
\newcommand{\flechedroite}[1]{
  \settowidth{\unitlength}{\mbox{$#1$}}
  \settoheight{\Taille}{\mbox{${\scriptstyle #1}$}}
  \addtolength{\Taille}{1ex}
  \addtolength{\unitlength}{4ex}
  \raisebox{0.5ex}{
  \begin{picture}(1,0)
    \put(0,0){\vector(1,0){1}}
    \put(0.5,0){\makebox(0,0){${\scriptstyle #1}$ \vspace{\the\Taille}}}
  \end{picture}}}
\newcommand{\flechegauche}[1]{
  \settowidth{\unitlength}{\mbox{$#1$}}
  \settoheight{\Taille}{\mbox{${\scriptstyle #1}$}}
  \addtolength{\Taille}{1ex}
  \addtolength{\unitlength}{4ex}
  \raisebox{0.5ex}{
  \begin{picture}(1,0)
    \put(1,0){\vector(-1,0){1}}
    \put(0.5,0){\makebox(0,0){${\scriptstyle #1}$ \vspace{\the\Taille}}}
  \end{picture}}}

\newtheorem{definition}{Definition}[section]

\newtheorem{proposition}{Proposition}[section]

\begin{document}
%\begin{flushright}
%{\it \bf ICMPA-MPA/2011}
%\end{flushright}
%\begin{titlepage}
\begin{center}
{\Large \bf {Spectrum
 of the harmonic oscillator in a general noncommutative phase space}}

Mahouton Norbert Hounkonnou$^{1,\dag}$ and Dine Ousmane
Samary$^{1,*}$

 $^{1}${\em University of Abomey-Calavi,\\
International Chair in Mathematical Physics
and Applications}\\
{\em (ICMPA--UNESCO Chair), 072 B.P. 50  Cotonou, Republic of Benin}\\

E-mails:   $^{\dag}$norbert.hounkonnou@cipma.uac.bj,\\
$^{*}$ousmanesamarydine@yahoo.fr.

%\maketitle
\begin{abstract}
Harmonic oscillator, in 2-dimensional noncommutative phase space with non-vanishing momentum-momentum
commutators, is studied using an algebraic approach. The corresponding
eigenvalue problem is solved and discussed.
%The physical solution is satisfied by certains
%relations under the deformation parameters.
\end{abstract}
\end{center}

\vspace{1cm}
%\end{titlepage}

%\section{Introduction}

In recent years, there is an increasing interest in the application of noncommutative (NC) geometry
 to  physical problems \cite{Connes} in solid-state and particle physics \cite{seiberg}, mainly motivated by  the idea of   a strong
connection of noncommutativity with   field and string theories.  Besides, the evidence coming from the latter and other approaches to
the issues of quantum gravity suggests that attempts to unify gravity and quantum mechanics
could ultimately lead to a non-commutative geometry of spacetime. The
phase space of ordinary quantum mechanics is a well-known example of noncommuting space \cite{Weyl}.
 The momenta of a system in the presence of a magnetic field are
noncommuting operators as well.
Since the noncommutativity between spatial and time coordinates may lead to some problems with unitarity and causality,
 usually only spatial noncommutativity is considered.
Besides, so far quantum theory on the NC space has been extensively studied,
the main approach is based on the Weyl-Moyal correspondence which amounts to replacing the usual product by a
$\star-$product in the NC space.
Therefore, deformation quantization has special significance in the study of physical
systems on the NC space. Moreover, the problem of quantum mechanics on NC
spaces can be understood in the framework of deformation
quantization \cite{Bayen}-\cite{Hir}. In the same vein,
some works on harmonic oscillators (ho) in the NC space from the point of view of deformation quantization have been reported in
\cite{Agapitos}-\cite{Land} and references therein.
In this paper, we consider different representations of a harmonic oscillator in a general full noncommutative phase space with
both the spatial and momentum coordinates being noncommutative.
 Indeed, noncommutativity between momenta arises naturally  as a consequence of noncommutativity between coordinates, as momenta
are defined to be the partial derivatives of the action with respect to the noncommutative coordinates. This work continues the investigations
pursued in \cite{Agapitos},\cite{Lin} and \cite{Lin2} devoted
to the study of a quantum exactly solvable $D$-dimensional NC oscillator with quasi-harmonic
behavior. We intend to extend previous results presenting a similar analysis to the quantum
version of the two-dimensional NC system with non-vanishing momentum components. For additional details in the motivation, see \cite{Agapitos}.
The physical model resembles to the Landau problem in NC quantum mechanics extensively studied  in the literature. See \cite{Polykronakos} and \cite{Larisa}
and references therein for more details. Broadly put, it is worth noticing that the description of a system of a
two-dimensional harmonic oscillator in a full NC phase space
is equivalent to that of the model of a two-dimensional ho in a constant
magnetic field in some NC space.

Consider a 2$D$ general NC phase space.
The coordinates of position and momentum,
  $x=(x^1 , x^2)$ and $p=(p^1 , p^2 ),$ modeling the classical system of a two-dimensional ho
maps into  their respective quantum
operators $\hat{x}$ and $\hat{p}$ giving rise to the Hamiltonian operator
\begin{eqnarray}\label{hamil}
\hat{H}=\frac{1}{2}\Big(\hat{p}_\mu\hat{p}^\mu+\hat{x}_\mu\hat{x}^\mu\Big)
\end{eqnarray}
with commutation relations
\begin{eqnarray}\label{alge}
[\hat{x}^\mu,\hat{p}^\nu]=i\hbar_{eff}\delta^{\mu\nu},\,\,\,  [\hat{x}^\mu,\hat{x}^\nu]=i\Theta^{\mu\nu},\,\,\,
[\hat{p}^\mu,\hat{p}^\nu]=i\bar{\Theta}^{\mu\nu},\quad \mu,\nu=1,2
\end{eqnarray}
where  $\Theta^{\mu\nu}$ and  $\bar{\Theta}^{\mu\nu}$ are skew-symmetric tensors carrying the dimensions of (length)$^2$
and (momentum)$^2,$ respectively.
The effective Planck constant
is given by
\begin{eqnarray}
\hbar_{eff}=\hbar\Big(1+\frac{\Theta^{\mu\nu}\bar{\Theta}^{\mu\nu}}{4D\hbar^2}\Big),
\end{eqnarray}
where $D=2$ is the dimension of the NC space.
One can readily check that one can rewrite the operators $\hat{x}^\mu$ and $\hat{p}^\nu$ as
\begin{eqnarray}\label{rrr}
\hat{p}^\mu=\hat{\pi}^\mu+\frac{1}{2\hbar}\bar{\Theta}^{\mu\nu}\hat{q}_\nu,\quad
\hat{x}^\mu=\hat{q}^\mu-\frac{1}{2\hbar}\Theta^{\mu\nu}\hat{\pi}_\nu
\end{eqnarray}
in terms of  $\hat{\pi}^\mu$ and $\hat{q}^\nu$ that obey the standard Weyl-Heisenberg algebra
\begin{eqnarray}\label{ctran}
[\hat{q}^\mu, \hat{\pi}^\nu]=i\hbar\delta^{\mu\nu};\quad [\hat{q}^\mu,\hat{q}^\nu]=0=[\hat{\pi}^\mu,\hat{\pi}^\nu].
\end{eqnarray}

%\section{Noncommutative Formulation in Relevant representations}
%In this section we shall be concerned with quantum systems
%whose dynamics is described by a self-adjoint Hamiltonian
%$\hat{H}(x,p)$, made up of the Cartesian coordinates $x=(\hat{x}^1, \hat{x}^2)$
%and their canonically conjugate momenta $p=(\hat{p}^1, \hat{p}^2)$.
%However, unlike the usual case, coordinates and momenta are
%supposed to obey the commutation
%rules (\ref{alge}).
In the deformation quantization theory of a classical system in the noncommutative
space, one treats $(x, p)$ and their functions as classical quantities, but replaces the ordinary
product between these functions by the following generalized
$\star-$product
\begin{eqnarray}
\star&=&\star_{\hbar_{eff}}\star_{\Theta}\star_{\bar{\Theta}} \cr
&=&\exp\Big[\frac{i\hbar_{eff}}{2}\Big(\overleftarrow{\partial}_{x^\mu}\overrightarrow{\partial}_{p^\mu}
-\overleftarrow{\partial}_{p^\mu}\overrightarrow{\partial}_{x^\mu}\Big)+\frac{i\Theta^{\mu\nu}}{2}
\overleftarrow{\partial}_{x^\mu}\overrightarrow{\partial}_{x^\nu}
%\cr
%&&
+\frac{i\bar{\Theta}^{\mu\nu}}{2}
\overleftarrow{\partial}_{p^\mu}\overrightarrow{\partial}_{p^\nu}\Big].
\end{eqnarray}
The variables $x^\mu$, $p^\mu$ on the NC phase
space satisfy the following commutation relations similar to (\ref{alge})
\begin{eqnarray}\label{alge}
&[x^\mu,p^\nu]_\star=i\hbar_{eff}\delta^{\mu\nu},\,\,\,  [x^\mu,x^\nu]_\star=i\Theta^{\mu\nu},\,\,\,
[p^\mu,p^\nu]_\star=i\bar{\Theta}^{\mu\nu}\\
&\mu, \nu= 1,2\nonumber
\end{eqnarray}
 with the following uncertainty relations
\begin{eqnarray}
\Delta x^1 \Delta x^2\geqslant\frac{\Theta}{2} \qquad \Delta p^1 \Delta p^2\geqslant\frac{\bar{\Theta}}{2}
\cr \Delta x^1 \Delta p^1\geqslant\frac{\hbar_{eff}}{2} \quad \Delta x^2 \Delta p^2\geqslant\frac{\hbar_{eff}}{2}.
\end{eqnarray}
The first two uncertainty relations show that measurements of positions and momenta in
both directions $x^1$ and $x^2$ are not independent. Taking into account the fact that $\Theta$ and $\bar{\Theta}$
have dimensions of (length)$^2$ and (momentum)$^2$ respectively, then $\sqrt{\Theta}$ and $\sqrt{\bar{\Theta}}$ define
fundamental scales of length and momentum which characterize the minimum uncertainties
possible to achieve in measuring these quantities.  One expects these fundamental scales to
be related to the scale of the underlying field theory (possible the string scale), and thus to
appear as small corrections at the low-energy level or quantum mechanics.
Commonly, the time evolution function for a time-independent Hamiltonian $H$ of a system is
denoted by the $\star-$exponential function denoted here by $e^{(.)}_\star:$
\begin{eqnarray}
e_\star^{\frac{Ht}{i\hbar_{eff}}}:=\sum_{n=0}^{\infty}\frac{1}{n!}\Big(\frac{t}{i\hbar_{eff}}\Big)^n
\overbrace{H\star H\star\cdots\star H}^{\mbox{n times}},
\end{eqnarray}
which is the solution of the following time-dependent  Schrodinger equation
\begin{eqnarray}
i\hbar_{eff}\frac{d}{dx}e_\star^{\frac{Ht}{i\hbar_{eff}}}&=&H(x,p)\star e_\star^{\frac{Ht}{i\hbar_{eff}}}\cr
&=&H\Big(x^\mu+\frac{i\hbar_{eff}}{2}\partial_{p^\mu}+\frac{i\Theta^{\mu\rho}}{2}\partial_{x^\rho},
p^\nu-\frac{i\hbar_{eff}}{2}\partial_{x^\nu}
%\cr
%&&
+\frac{i\bar{\Theta}^{\mu\sigma}}{2}\partial_{x^\sigma}\Big) e_\star^{\frac{Ht}{i\hbar_{eff}}}.
\end{eqnarray}
There corresponds the generalized $\star-$eigenvalue time-independent Schrodinger equation:
\begin{eqnarray}\label{6}
H\star \mathcal{W}_n=\mathcal{W}_n\star H= \mathcal{E}_n \mathcal{W}_n
\end{eqnarray}
where $\mathcal{W}_n$ and $\mathcal{E}_n$ stand for  the Wigner function and the corresponding energy eigenvalue of the system.
The Fourier-Dirichlet expansion for the time-evolution function defined as
\begin{eqnarray}
 e_\star^{\frac{Ht}{i\hbar_{eff}}}=\sum_{n=0}^\infty e^{\frac{-i\mathcal{E}_n t}{\hbar_{eff}}}\mathcal{W}_n
\end{eqnarray}
links the Wigner function to the $\star-$exponential function.
%\subsection{Quick review of ho  in NC  $(q, \pi)-$representation}
Provided the above, the operators on a NC Hilbert space can be represented by the functions on a NC phase space, where the operator product is
replaced by relevant star-product. The algebra of functions of such noncommuting coordinates can be replaced
by the algebra of functions on ordinary spacetime, equipped with a NC star-product.
So, considering the transformations (\ref{rrr}) and leaving out the operator symbol $\hat{}$, we arrive
at $(q,\pi)$ phase space and the commutation relation change
into (\ref{ctran}),
 with the star-product defined in the following way.
\begin{definition}
Let $C^\infty(\mathbb{R}^4)$ be the space of smooth functions $f: \mathbb{R}^4\rightarrow \mathbb{C}$. For
$f,g\in C^\infty(\mathbb{R}^4)$, the formal star product is defined by
\begin{eqnarray}
f\star g=f \exp\Big[\frac{i\hbar}{2}
\overleftarrow{\partial}_\mu J^{\mu\nu}\overrightarrow{\partial}_\nu\Big]g.
\end{eqnarray}
Here the smooth functions $f$ and $g$   depend on the real variables $q^1$, $q^2$, $\pi^1$ and $\pi^2$,
and
\begin{eqnarray}
\overleftarrow{\partial}_\mu J^{\mu\nu}\overrightarrow{\partial}_\nu&=&
\left(\frac{\overleftarrow{\partial}}{\partial q^1},
\frac{\overleftarrow{\partial}}{\partial \pi^1},\frac{\overleftarrow{\partial}}{\partial q^2},
\frac{\overleftarrow{\partial}}{\partial \pi^2}\right)\left(\begin{array}{cccc}
 0&1&0&0\\
-1&0&0&0\\
0&0&0&1\\
0&0&-1&0
                                                                      \end{array}
\right)\left(\begin{array}{cccc}
              \frac{\overrightarrow{\partial}}{\partial q^1}\\ \frac{\overrightarrow{\partial}}{\partial \pi^1}\\
\frac{\overrightarrow{\partial}}{\partial q^2}\\
\frac{\overrightarrow{\partial}}{\partial \pi^2}
             \end{array}
\right)\cr
&=&\frac{\overleftarrow{\partial}}{\partial q^1}\frac{\overrightarrow{\partial}}{\partial \pi^1}
-\frac{\overleftarrow{\partial}}{\partial \pi^1}\frac{\overrightarrow{\partial}}{\partial q^1}
+\frac{\overleftarrow{\partial}}{\partial q^2}\frac{\overrightarrow{\partial}}{\partial \pi^2}
-\frac{\overleftarrow{\partial}}{\partial \pi^2}\frac{\overrightarrow{\partial}}{\partial q^2}.
\end{eqnarray}
\end{definition}
Therefore,  the star product $f\star g$ represents a deformation of the classical product $fg$. This deformation depends
on the Planck constant $\hbar$. In term of physics, the difference $f\star g -fg$ describes quantum fluctuation depending
on $\hbar$. For the present case,
\begin{eqnarray}\label{idnew}
& q^\mu\star\pi^\nu-q^\mu\pi^\nu=\frac{i\hbar}{2}\delta^{\mu\nu},\quad \pi^\nu\star q^\mu-
\pi^\nu q^\mu=-\frac{i\hbar}{2}\delta^{\mu\nu}.\quad\mbox{ Hence }\cr
& [q^\mu,\pi^\nu]_\star=i\hbar\delta^{\mu\nu}.
\end{eqnarray}
Building  now, in  the standard manner, the creation and annihilation operators of ho system as
\begin{eqnarray}
a_l=\frac{q^l+i\pi^l}{\sqrt{2}}\quad \quad \bar{a}_l= \frac{q^l-i\pi^l}{\sqrt{2}}\quad l=1,2
\end{eqnarray}
and using the polar coordinates  such that
\begin{eqnarray}\label{polar}
q^l=\rho_l \cos\varphi_l,\qquad \pi^l= \rho_l \sin\varphi_l,
\end{eqnarray}
we solve the right and left eigenvalue equations
\begin{eqnarray}
a_l\star f_{mn}=\sqrt{m\hbar}f_{m-1,n} \quad \bar{a}_l\star f_{mn}=\sqrt{(m+1)\hbar}f_{m+1,n}\cr
f_{mn}\star a_l=\sqrt{(n+1)\hbar}f_{m,n+1}\quad f_{mn}\star \bar{a}_l=\sqrt{n\hbar} f_{m,n-1}
\end{eqnarray}
to find the eigenfunctions $f_{mn}$ as
\begin{eqnarray}\label{base}
f_{mn}\equiv2(-1)^{m}\sqrt{\frac{m!}{n!}}e^{i(n-m)\varphi_l}\Big(\frac{2\rho_l^{2}}{\hbar}
\Big)^{\frac{n-m}{2}}L_{m}^{n-m}\Big(\frac{2\rho_l^{2}}{\hbar}\Big)
e^{-\frac{\rho_l^{2}}{\hbar}},\quad m,n\in\mathbb{N}
\end{eqnarray}
with
\begin{eqnarray}
f_{00}=2 e^{-\rho_l^2/\hbar}.
\end{eqnarray}
  $L_{m}^{n-m}$ are the generalized Laguerre polynomials defined for $n=0,1,2,\cdots,\,\, \alpha>1,$ by
\begin{eqnarray}
L_n^\alpha(x)=\frac{1}{n!}e^x x^{-\alpha}\frac{d^n}{dx^n}(x^{n+\alpha}e^{-x}) =\sum_{k=0}^n
\frac{\Gamma(n+\alpha+1)}{\Gamma(k+\alpha+1)}\frac{(-x)^k}{k!(n-k)!}.
\end{eqnarray}
Then the states defined by $b_{mn}^{(4)}=f_{m_1n_1}f_{m_2n_2},$ where $m=(m_1,m_2)$, $n=(n_1,n_2)$, $m_1, m_2, n_1, n_2\in\mathbb{N},$
exactly solve the right and left eigenvalue problems of the Hamiltonian $H_0=\sum_{l=1}^2 \bar{a}_la_l$ as
\begin{eqnarray}\label{pti}
H_0\star b_{mn}^{(4)}=\hbar(|m|+1) b_{mn}^{(4)}\quad \mbox{ and }\quad b_{mn}^{(4)}\star
H_0= \hbar(|n|+1)b_{mn}^{(4)}
\end{eqnarray}
where $|m|=m_1+m_2$.

%\subsection{Ho in NC $(x,p)-$representation versus $(q,\pi)$}
Now, consider  the Hamiltonian
(\ref{hamil}) and use the relation (\ref{ctran}) to re-express it with the help of variables $q$ and $\pi$ as follows:
\begin{eqnarray}\label{gh}
H=H_0+H_L+ H_q(\bar{\Theta})+ H_\pi(\Theta)
\end{eqnarray}
where
\begin{eqnarray}
H_0=\frac{1}{2}\Big((q^1)^2 +(q^2)^2 +(\pi^1)^2 +(\pi^2)^2 \Big)
\end{eqnarray}
\begin{eqnarray}
H_L=-\frac{\Theta+\bar{\Theta}}{2\hbar} \overrightarrow{ q}\wedge \overrightarrow{ \pi}\qquad
\overrightarrow{ q}\wedge \overrightarrow{ \pi}=q^1\pi_2-q^2\pi_1
\end{eqnarray}
and
\begin{eqnarray}
H_q(\bar{\Theta})=\frac{\bar{\Theta}^2}{8\hbar^2}\Big((q^1)^2 +(q^2)^2 \Big)\quad
H_\pi(\Theta)=\frac{\Theta^2}{8\hbar^2}\Big((\pi^1)^2 +(\pi^2)^2 \Big).
\end{eqnarray}
It is a matter of computation to verify that the Hamiltonians $H_0$
  and $H_L$  $\star-$commute. Idem for    the Hamiltonians $H_L$
  and
 $H_I=H_q(\bar{\Theta})+ H_\pi(\Theta).$ Therefore,   the  Hamiltonians of family $\{H_0,H_L\},$ (respectively
$\{H_L, H_I\}$) can be simultaneously measured.

\begin{proposition}
 In the case  $\Theta=-\bar{\Theta}$, the Hamiltonian $H$ can be expressed as
\begin{eqnarray}
H=\Big(1+\frac{\Theta^2}{4\hbar^2}\Big)H_0
\end{eqnarray}
and the states $b_{mn}^{(4)}$ solve the right and left eigenvalue problems of H as
\begin{eqnarray}
H\star b^{(4)}_{mn}=\mathcal{E}_{m0}^R b^{(4)}_{mn}
\quad\quad \mathcal{E}_{m0}^R=\hbar \Big(1+\frac{\Theta^2}{4\hbar^2}\Big)(|m|+1)
\end{eqnarray}
and
\begin{eqnarray}
 b^{(4)}_{mn}\star H=\mathcal{E}_{0n}^L b^{(4)}_{mn}
\quad\quad \mathcal{E}_{0n}^L=\hbar \Big(1+\frac{\Theta^2}{4\hbar^2}\Big)(|n|+1)
\end{eqnarray}
where $m=(m_1,m_2)$, $n=(n_1,n_2)$ $m_1, m_2, n_1, n_2\in \mathbb{N}$, $|m|=m_1+m_2.$
\end{proposition}

In the general  $(q,\pi)-$representation,
the problem to solve is  equivalent to that of the two-dimensional Landau problem in a symmetric gauge
on a noncommutative space. Indeed, the Hamiltonian H can be re-transcribed as
\begin{eqnarray}\label{gh1}
 H=\frac{\alpha^2}{2}\Big((q^1)^2 +(q^2)^2 \Big)+\frac{\beta^2}{2}\Big((\pi^1)^2 +(\pi^2)^2  \Big)-\gamma
\overrightarrow{ q}\wedge \overrightarrow{ \pi}=:H^\natural_0+ H_L
\end{eqnarray}
where
\begin{eqnarray}
 \alpha^2=1+\frac{\bar{\Theta}^2}{4\hbar^2},\qquad \beta^2=1+\frac{\Theta^2}{4\hbar^2},\qquad \gamma=
\frac{\Theta+\bar{\Theta}}{2\hbar}
\end{eqnarray}
Remark  that the Hamiltonian terms $H^\natural_0$ and  $H_L$ commute. Therefore, the eigenvectors of
$\{H^\natural_0, H_L\}$ are automatically  eigenvectors of $H$. As matter of convenience,  to solve the Schr\"odinger eigen-equation,
 let us choose the polar coordinates
\begin{eqnarray}
q^1=\rho\cos\varphi\qquad q^2= \rho\sin\varphi
\end{eqnarray}
and assume the variable separability to write
\begin{eqnarray}
\tilde{f}(\rho,\varphi)=\xi(\rho)e^{ik\varphi},\quad k=0,\pm 1,\pm 2,\cdots
\end{eqnarray}
Then, from the static Schr\"odinger equation on NC space,
 $H\star \tilde{f}(\rho,\varphi)= \mathcal{E}
\tilde{f}(\rho,\varphi),$
 we deduce the radial equation as
follows:
\begin{eqnarray}
\Big[-\frac{\hbar^2\beta^2}{2}\Big(\frac{\partial^2}{\partial\rho^2}+\frac{1}{\rho}
\frac{\partial}{\partial\rho}\Big)+\frac{\alpha^2}{2}\rho^2 -\gamma\hbar k \Big]\xi(\rho,\varphi)=
\mathcal{E}\xi(\rho,\varphi)
\end{eqnarray}
yielding the spectrum of $H$ under the form
\begin{eqnarray}
\mathcal{E}=\hbar\frac{\alpha^2}{\beta^2}(n+1)
-\hbar\gamma k, \quad n=0,1,2,\cdots
\end{eqnarray}
with
\begin{eqnarray}
\xi(\rho, \varphi)\propto e^{-\frac{\alpha}{\hbar\beta}\rho^2}H_{n}\left(\frac{\alpha}{\hbar\beta}\rho^2\right).
\end{eqnarray}
The last term of the energy spectrum $\mathcal{E}$ falls down when $\gamma= 0,$ i.e.
$\Theta= -\bar{\Theta}.$ In this case, $ \alpha^2=  \beta^2$ and we recover the discrete spectrum of the usual
two-dimensional harmonic oscillator as expected.
%\section{Conclusion and Remark}
The results obtained here  can be reduced to specific expressions reported in the
literature \cite{Agapitos} for particular cases. Besides, the formalism displayed in this work permits to avoid
the appearance of infinite degeneracy of states observed when $\hbar_{eff}^2-\Theta\bar{\Theta}=0$ in
 \cite{Polykronakos}  where the phase space is divided  into two phases based on the following conditions on the deformation parameters:
\begin{itemize}
 \item  Phase I for $\hbar_{eff}^2-\Theta\bar{\Theta}>0$
\item  Phase II for $\hbar_{eff}^2-\Theta\bar{\Theta}<0.$
\end{itemize}
%and it critical point  correspond to reduction of dimensions in phase space
%and to infinite degeneracy of states and is related to the NC Landau problem.
Finally, let us mention that  the direct computation of the energy spectrum from the relation (\ref{gh}) instead of (\ref{gh1}) introduces an unexpected
feature, i.e.  the energy spectrum depends on the phase space variables as it should not be with respect to the study performed
 in \cite{Larisa}. Such a pathology is generated by the phase space variable dependence of the commutator
\begin{eqnarray}
 [H_0, H_I]_\star=i\frac{\Theta^2-\bar{\Theta}^2}{4\hbar}(q^1\pi^1+q^2\pi^2).
\end{eqnarray}
This could explain why previous investigations (see \cite{Agapitos}, \cite{Wei}
and \cite{Sayidjamal} and references therein) were restricted to the cases $\Theta= \pm\bar{\Theta}.$
\section*{Acknowledgements}
This work is partially supported by the ICTP through the
OEA-ICMPA-Prj-15. The ICMPA is in partnership with the Daniel
Iagolnitzer Foundation (DIF), France.

\end{document}